\documentclass[manuscript,12pt]{aastex} 

\newcommand{\degree}{^{\circ}}

\newcommand{\eg}{{\it e.g.\ }}
\newcommand{\ie}{{\it i.e.\ }}

\newcommand{\vex}{v_{+}}

\def\las{\mathrel{\hbox{\rlap{\hbox{\lower3pt\hbox{$\sim$}}}\hbox{\raise2pt\hbox{$<$}}}}}
\def\gas{\mathrel{\hbox{\rlap{\hbox{\lower3pt\hbox{$\sim$}}}\hbox{\raise2pt\hbox{$>$}}}}}
\def\dg{{}^\g\setbox\help=\hbox{${}^\g$}\setbox\punt=\hbox{$.$}\skip
\terug=-.5\wd\help plus0pt minus0pt\advance\skip\terug by -0.17em plus0em
minus0em\hskip\skip\terug\skip\vooruit=-\skip\terug\advance\skip\vooruit by
-\wd\punt.\hskip\skip\vooruit}

\author{Paul Wiegert}
\affil{Dept. of Physics and Astronomy\\The University of Western Ontario London Canada\\Submitted to Icarus April 8 2014}

\title{Hyperbolic meteors: interstellar or generated locally via the gravitational slingshot effect?}
\shorttitle{Hyperbolic meteoroids}

\begin{document}		% Critical

%% Next comes the abstract, notice the curly-braces surrounding the
%% text.

\begin{abstract}

The arrival of solid particles from outside our solar system would
present us with an invaluable source of scientific
information. Attempts to detect such interstellar particles among the
meteors observed in Earth's atmosphere have almost exclusively assumed
that those particles moving above the Solar System's escape speed --
particles on orbits hyperbolic with respect to the Sun-- were
precisely the extrasolar particles being searched for.  Here we show
that hyperbolic particles can be generated entirely within the Solar
System by gravitational scattering of interplanetary dust and
meteoroids by the planets. These particles have necessarily short
lifetimes as they quickly escape our star system; nonetheless some may
arrive at Earth at speeds comparable to those expected of interstellar
meteoroids. Some of these are associated with the encounter of planets
with the debris streams of individual comets: Comet C/1995 O1
Hale-Bopp's 1996 pre-perihelion encounter with Jupiter could have
scattered particles that would have reached our planet with velocities
of almost 1 km~s$^{-1}$ above the hyperbolic velocity at Earth;
however, such encounters are relatively rare. The rates of occurrence
of hyperbolically-scattered sporadic meteors are also quite low. Only
one of every $\sim10^{4}$ optical meteors observed at Earth is
expected to be such a locally generated hyperbolic and its
heliocentric velocity is typically only a hundred meters per second
above the heliocentric escape velocity at Earth's orbit. The majority
of such gravitationally-scattered hyperbolics originate at Mercury,
though Venus and Mars also contribute. Mercury and Venus are predicted
to generate weak 'hyperbolic meteor showers': the restrictive geometry
of scattering to our planet means that a radiant near the Sun from
which hyperbolic meteors arrive at Earth should recur with the
planet's synodic period. However, though planetary scattering can
produce meteoroids with speeds comparable to interstellar meteors and
at fluxes near current upper limits for such events, the majority of
this locally-generated component of hyperbolic meteoroids is just
above the heliocentric escape velocity and should be easily
distinguishable from true interstellar meteoroids.

\end{abstract}


%\notetoeditor{Figures: 8   Tables: 1\\
%CORRESPONDING AUTHOR: Paul Wiegert\\Dept. of Physics and Astronomy, The
%University of Western Ontario\\London Ontario, Canada N6A 3K7\\pwiegert@uwo.ca
%\\Phone/fax: 1-519-661-2111x81327/1-519-661-2033
%}

\maketitle			% Use the \author, \title and \date info

\section{Introduction}

The first measurement of a meteor velocity may have been due to
\cite{elk00}. He used a bicycle wheel as the basis for a rotating
shutter that would interrupt the meteor's image on a photograph, and
the segmented image was used to determine the meteor's velocity. The
idea of using photographs to measure meteor velocity goes back
further, at least to \cite{lan60} but was not initially
widely-used. The difficulty with photographic observations was its
limited sensitivity in its early days: a hundred hours of observation
might be required for a single successful result \eg
\cite{lov54}. Though photography goes back to the early 1800's, as
late as 1932 \cite{shaopiboo32} noted that ``several hundred meteors
are visible to the unaided eye to one that can be photographed''.

Naked-eye visual observations provided the first substantial number of
meteor velocity measurements, along with the first indication of
meteors that might be from outside our Solar System.  Von Nei{\ss}l
and Hoffmeister's Fireball Catalogue \citep{vonhof25} contains
visually determined orbits of fireballs. Many of their entries are hyperbolic
with respect to the Sun, that is, their velocities are so large that
they cannot be gravitationally bound to our solar system. At the
Earth's orbit, the parabolic or escape velocity with respect to the
Sun is about 42 km~s$^{-1}$, and 79\% of \cite{vonhof25}'s orbits exceed this
value, some ranging up to 99 km~s$^{-1}$ (as quoted in \cite{lov54}). 

A simple interpretation of hyperbolic meteors was that, since they
were not bound to our Solar System, they must be from outside it and
thus represent material originating elsewhere in our Galaxy. However,
not all researchers agreed that substantial numbers of meteors had
hyperbolic velocities, attributing them rather to measurement
error. The discussion of the reality of hyperbolic meteors centred on
the sporadic meteors: many showers were already accepted to be on
bound orbits near those of their parent comets and thus part of
streams of particles originating within our planetary system.  A
vigorous debate as to the existence of hyperbolic meteors spanned the
next few decades. \cite{fis28} and \cite{wat39} concluded that since
the hyperbolic meteors of \cite{vonhof25} largely coincided in time
with the major showers and in space with the ecliptic plane, that they
were unlikely to be of true interstellar origin; they instead
concluded that a systematic over-estimation of the velocities, which
were at the time still measured by observers using the naked eye, was
more likely. Others argued conversely that, if some showers were in
fact interstellar in nature, meteor showers and hyperbolic meteors
might be expected to coincide in some cases.

The problem was considered sufficiently important that the Harvard
Observatory organized the Arizona Expedition to resolve the question
\citep{shaopiboo32}. Ernst {\"O}pik led a campaign that erected two
'meteor houses' in Arizona where observers would record meteor data --
still taken visually by human observers -- in an organized fashion. The
houses, really small protective shelters for the observers, had
windows with built-in reticules to aid in positional measurements. The
campaign also made use of the clever 'double-pendulum' or 'rocking
mirror' technique whereby the meteors' motion would by translated into
a pseudo-cycloidal motion, the number of cusps/loops of which could be
used to facilitate trail length and speed measurements
\citep{shaopiboo32,mcfash10}.

The initial results of this work \citep{opi40} reported 57.2\% of
meteors as being hyperbolic, with heliocentric velocities in rare
cases reaching over 280 km~s$^{-1}$. The expedition leader
was aware of the potential pitfalls: ``As in all kinds
of visual observations of meteors in which the observer has finally to
rely upon his memory, considerable accidental and systematic errors
are involved in the observed velocities too; in a statistical
discussion of velocities such as given below the data must be freed,
in the first place, from the influence of these errors. Only after
that can the bearing of the statistical data upon cosmic problems be
investigated'' \citep{opi40}.

Despite the Arizona results, some astronomers remained sceptical that
sporadic meteors were interstellar. \cite{por43,por44}, working from
other visual observations, concluded that meteors are not hyperbolic
in any great numbers, and emphasized the need for a careful
statistical analysis of a sample with known errors. The interstellar
hypothesis received a serious blow when the first photographic meteor
studies \citep{whi40} identified the Taurid stream, once conjectured
to be an interstellar stream, and found it to be bound to the Sun and
associated with Comet Encke.

%  The interesting result of Lovell p 154 (Pultusk
%  meteorite). cite{jonsar84} gives a reference to \cite{gal68} for
%  this meteorite. \citep{wyl40} provides a discussion which concludes
%  it is not interstellar. There is a translation of Galle (who was a
%  co-discoverer of Neputne) result in \cite{gal68b}

Though {\"O}pik continued to stand by the Arizona expedition's
findings of a high fraction of hyperbolic sporadic meteors
\citep{opi50}, new photographic programs began finding the
interstellar fraction of meteors to be quite small.  The Harvard
Super-Schmidt photographic program \citep{jacwhi61} detected very few
hyperbolic orbits.  Radar meteor observations from Jodrell Bank
(\cite{almdavlov51,almdavlov52,almdavlov53,cle52}; see \cite{gun05}
for a review) and from Ottawa \citep{mck49,mck51} showed little or no
evidence for interstellar velocities. \cite{opi69} eventually
conceded that there was a failure in the basic assumptions underlying
the rocking mirror technique, due partly to height differences
between sporadic and shower meteors, and partly due to 'psychological'
differences in their perception by observers.

%Interstellar meteoric particles were reported by Pioneer 8 according
%to \citep{opi73b}, who references Berg at NASA. 

Though the hyperbolic component is now recognized to be small at
visual meteor sizes ($\gas$~1~mm), they have been detected
convincingly in interplanetary space at smaller sizes. Dust detectors
aboard the Ulysses, Galileo and Helios spacecraft
\cite[]{gruzoobag93,fridorgei99,krulanalt07} have detected very small
($10^{-18}-10^{-13}$~kg) grains moving at speeds above the local solar
system escape velocity and parallel to the local flow of interstellar
gas. This result provides perhaps the first generally-accepted
detection of interstellar meteoroids. However, these particles are too
small to be detected as meteors at the Earth: sizes $\gas10^{-10}$~kg
may be required for this.

Meteor radars are typically more sensitive that the human eye or
photographs and can detect particles much smaller than those seen in
early surveys of sporadic meteors.  The Advanced Meteor Orbit Radar
(AMOR) reported that a few percent of meteors with sizes of
$\sim50\mu$m or masses$\sim10^{-10}$~kg were hyperbolic, many of which were
proposed to be particles ejected from the $\beta$ Pictoris dust disk
\citep{bag99,bag00,baggal01}. Radar hyperbolics have also seen by
\cite{janmeimat01} at Arecibo though these were determined to be meteoroids
accelerated by solar radiation pressure and originating within our own
Solar System. Some Arecibo radar meteor detections have been
interpreted as true interstellar meteoroids
\citep{meijanmat02a,meijanmat02b}, and thus the possibility remains that
substantial numbers of interstellar meteoroids reach the Earth at
sizes that do not produce naked eye meteors.  However, not all
radar studies show evidence for hyperbolic meteors: radar observations
at the Canadian Meteor Orbit Radar (CMOR) do not contain appreciable
hyperbolics: less than 0.0008\% \citep{werbro04}.
 
Harvard super-Schmidt photographic observations \citep{mcrpos61},
photographic meteor observations \citep{babkra67}, TV meteor
observations \citep{jonsar85}, and image-intensified video meteor
studies \citep{hawwoo97} all show a minority fraction of hyperbolic
orbits. Between 1\% and 22\% of of meteors observed at the Earth by
various surveys, optical and radar-based, have shown a hyperbolic
component according to reviews by \cite{hawclowoo99} and
\cite{bagmarclo07}. It remains unclear whether these represent true
hyperbolics or result from experimental uncertainties.
\cite{muswerbro12} present image-intensified optical results of a
small number (22 of 1739) possible interstellar meteoroids but
ultimately attribute these to measurement errors. Work by Hajdukov\'a
and her colleagues \cite[]{hajpau02,hajhaj06,hajpau07,haj08,haj11} has
shown that -- even with modern photographic and video techniques -- in
many cases hyperbolic meteors only appear so as the result of
measurement errors.

The true population of interstellar meteoroids within our Solar System
remains unknown. The most recent theoretical work on the expected
component of true interstellar meteoroids in our Solar System is
\cite{murweicap04}, but the question must ultimately be answered by
measurement. However, even given an unequivocal measurement of
hyperbolic velocity for a meteor, the question remains: did the
particle originate outside our solar system? Given that processes
within our Solar System might produce particles with high
velocities and that could ``contaminate'' our sample of interstellar
meteors, we need to understand the population of hyperbolic meteors
produced internally to our planetary system in order to tease the two
apart.

Here we address the question of whether or not hyperbolic meteoroids
could be produced within our own Solar System, in particular by the
gravitational slingshot effect. It has long been recognized that
planetary scattering must produce some hyperbolic meteors whose origin
is contained wholly within our Solar System
\citep{lov54,opi69}. Not that gravitational scattering is the only
mechanism by which they might be produced.  The ejection processes of
comets may inject meteoroids directly onto hyperbolic orbits. Small
($\las 1\mu$m) cometary particles may feel enough radiation pressure
to be unbound from the Sun independent of their ejection velocity from
their parent (the so-called beta meteoroids).  Collisional or
rotational breakup of meteoroids and subsequent radiation pressure
modification of their orbits  are also thought to contribute to this
population \citep{zoober75}. The magnetic fields of the planets Jupiter
and Saturn are also able to accelerate electrically charged grains
to escape speeds, as was measured by the Ulysses and Cassini spacecraft
\citep{gruzoobag93,kemsrahor05,flakruham11}.

Most these mechanisms only produce hyperbolic meteoroids if the
particles are small enough that radiation pressure can play a role in
accelerating them out of our Solar System, and are not effective for
larger particles.  Particles small enough to be ejected via radiation
pressure are not easily detected by Earth-based meteor sensors, and so
reports of high-speed meteors are unlikely to be of this origin.
Direct cometary ejection may also produce larger particles on
hyperbolic orbits without the action of radiation forces; however,
meteoroids of this type are likely to be part of a freshly deposited
meteoroid stream and unlikely to be mistaken for interstellar
meteoroids. Gravitational scattering is perhaps the only source of
meteoroids larger than a few microns in size which could be readily
confused with an interstellar influx of particles.

Here we examine the properties of gravitationally scattered meteoroids
produced within our Solar System. Though in some sense
``contaminants'' of the interstellar meteoroid sample, their intrinsic
properties are of interest as well for they come to us directly from
the vicinity of the planets and may thus carry important information
about these bodies and their environments. The techniques of meteor
velocity measurement continue to improve and it is only a question of
time before a substantial number of reliable hyperbolic meteors is
measured: an understanding of the flux of such meteors produced within
our Solar System is needed before the true interstellar component can
be separated from the local one.

\section{Methods} \label{model}

\subsection{Scattering of a meteoroid by a planet} \label{scattering}

This work expands on the preliminary results of \cite{wie11}.  The
scattering is calculated analytically by taking a patched conic
approximation to the meteoroid's trajectory past the scattering
planet. The meteoroid is taken to be on a bound heliocentric orbit
prior to its intersecting the planet's Hill sphere. From this point
onward, it is assumed to travel on a two-body orbit around the planet
until it leaves the Hill sphere once more, and then is taken to return
to a purely heliocentric orbit. We allow the meteoroid to approach the
planet from an arbitrary direction, and the scattering takes place in
three dimensions. Here we are most interested in those particles whose
post-scattering orbits are unbound with respect to the Sun and which
also subsequently pass close to the Earth.

The procedure for calculating the scattering is straightforward and
well-known \eg \cite{roy78} but is rarely presented in its full
three dimensional form and so we outline it here.

For simplicity, we assume that the planet orbits the Sun on a circular
orbit of radius $a$ and velocity
\begin{equation}
v_p= \sqrt{\frac{G M_{\sun}}{a}} \label{vp-helio}
\end{equation}
where $G$ is the usual gravitational constant and $M_{\sun}$ is the
Sun's mass.  The planet's Hill sphere has a radius $H$ of
\begin{equation}
H = a \left(\frac{M_p}{3M_{\sun}}\right)^{1/3} 
\end{equation}
where $M_p$ is the planet's mass. Inside the Hill sphere, the particle
is taken to move on a Keplerian two-body path around the planet. Upon
reaching the Hill sphere again (which is guaranteed by conservation of
energy: the meteoroid cannot become bound to the planet in this
approximation), the meteoroid proceeds onwards along a heliocentric
Keplerian orbit.

\begin{figure}
\epsscale{1.0}
\plotone{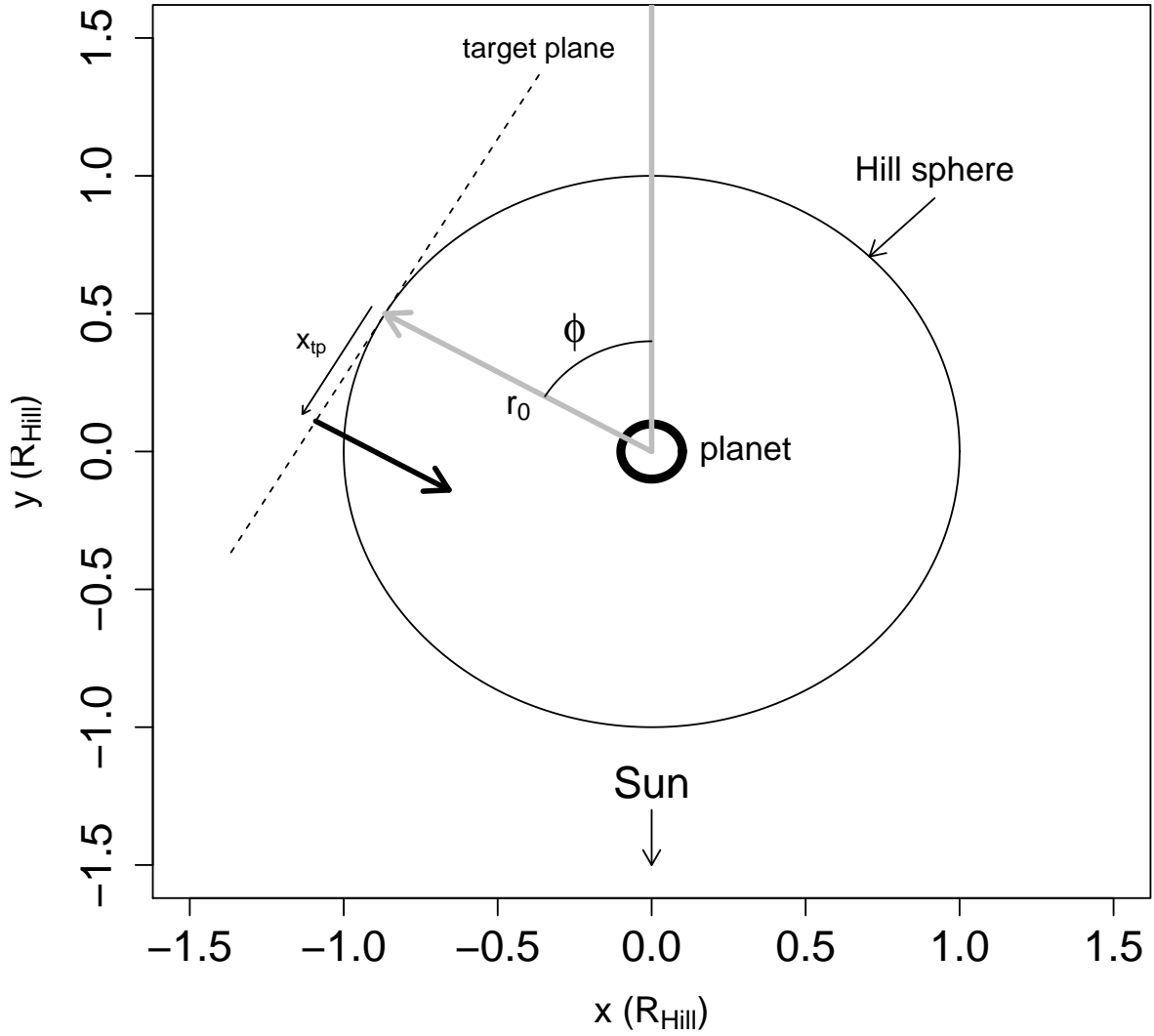}
\caption{The coordinate system used in this study. See the text for more details.}
\label{fi:gs-frame}
\end{figure}

The initial heliocentric approach velocity is taken to be
anti-parallel to the vector $\vec{r}_0$ which runs from the planet to
the centre of the target plane (see Figure~\ref{fi:gs-frame}). The
vector $\vec{r}_0$ is defined to be
\begin{equation}
\vec{r}_0 = H( -\sin \phi  \cos \psi, \cos \phi \cos \psi, \sin \psi )
\end{equation}
where $\phi \in [0,2 \pi)$ is the angle measured counterclockwise from
  the $y-z$ plane along the Hill sphere (see Fig.~\ref{fi:gs-frame})
  The latitude $ \psi \in [-\frac{\pi}{2}, \frac{\pi}{2}]$ is the
  angle between $\vec{r}_0$ and the planet's orbital plane, and is
  zero if the particle's motion is parallel to the ecliptic plane. The
  Sun-planet line defines the $y$ direction and the $z$-axis is
  perpendicular to the plane of the planet's orbit. The heliocentric
  velocity $\vec{v}_p$ of the planet is in the negative $x$-direction.

%\centerline{FIGURE CAPTIONS}
%Fig.~\ref{fi:gs-frame}: {d observations (black line) for all meteors}

A meteoroid enters the Hill sphere with a heliocentric velocity $\vec{v}_i$
anti-parallel to $\vec{r}_0$ given by
\begin{equation}
\vec{v}_i = v_i (  \sin \phi  \cos \psi, - \cos \phi \cos \psi, -\sin \psi ) \label{eq:vi-helio}
\end{equation}
The meteoroid crosses into the Hill sphere at the location ($x_{tp}$, $y_{tp}$) on the target plane where both $x_{tp}$ and $y_{tp}$ may run from $-H$ to $+H$. Note that ($x_{tp}$, $y_{tp}$) are defined to correspond to the $x$ and $y$ positions on the target plane as seen by the approaching particle (note the definition of $x_{tp}$ in Figure~\ref{fi:gs-frame}).

 Taking $x_{tp} \equiv H \sin \beta$ and $y_{tp} \equiv H \sin \zeta$ we can write the vector $\vec{R}_i$ from the planet to where the particle enters its Hill sphere as
\begin{equation}
\vec{R}_i = H( -\sin (\phi+\beta)  \cos( \psi+\zeta), \cos (\phi+\beta) \cos (\psi+\zeta), \sin (\psi+\zeta) ) 
\end{equation}
The initial planetocentric velocity $\vec{V}_i$ is then given by the vector sum of Eq.~\ref{eq:vi-helio} and $\vec{v}_p$
\begin{equation}
\vec{V}_i = ( v_i \sin \phi  \cos \psi + v_p, -v_i \cos \phi \cos \psi, -v_i \sin \psi )
\end{equation}
With the initial  planetocentric position $\vec{R}_i$ and velocity $\vec{V}_i$
now defined, we can proceed to calculate the parameters of the scattering.

A check of the sign of the dot product $\vec{R}_i \cdot \vec{V}_i$
allows a determination as to whether or not the particle is
approaching the planet: if this quantity is greater than zero, the
particle is receding in the planetocentric frame and no scattering
occurs. If the particle does approach the planet, a vector
along the pole $\vec{U}$ of the planetocentric orbit is given by the
vector cross-product
\begin{equation}
\vec{U} = \vec{R}_i \times \vec{V}_i
\end{equation}

As the particle passes its point of closest approach to the planet and
then recedes its initial velocity vector is rotated through an angle
$\gamma$. A well-known result of the scattering problem, in this case
$\gamma$ can be shown to be
\begin{equation}
\gamma = 2 \arctan \left( \frac{G M_p}{B |\vec{V}_i|^2} \right) \label{eq:gamma}
\end{equation}
where $B \ge 0$ is the impact parameter in the planetocentric frame 
\begin{equation}
B =  \frac{| \vec{U} |}{|\vec{V}_i|} 
\end{equation}
where the absolute value symbol indicates taking the length of the
vector in question.

At this point one may also wish to calculate the planetocentric eccentricity
\begin{equation}
e = \frac{1}{\sin( \gamma/2)} \label{eq:planetecc}
\end{equation}
Note that $e$ in Eqn.~\ref{eq:planetecc} cannot be less than unity,
since the sine function cannot exceed one, and so the particle's orbit
relative to the planet is always unbound. The closest approach
distance to the planet $q$ is given by
\begin{equation}
q = GM_p(e-1)/|\vec{V}_i|^2 
\end{equation}
which is valid for the case $e > 1$ which applies here.

The pericentre distance should be checked for collision with the
planet. The planet radius $r_{p}$ is here taken to be distance from
the centre of the planet where the atmospheric density drops to
$10^{-6}$~kg~ m$^{-3}$. This density occurs at an altitude of 100~km
on the Earth and corresponds to the start of meteoroid ablation.  If
$q < r_{p}$, the meteoroid is deemed destroyed by a collision with the
planet or by ablation within its atmosphere. The planetary data used
is listed in Table~\ref{ta:planetarydata}.

\begin{table}
\begin{center}
\begin{tabular}{lccccc}
 planet & $1/M_p (M_\sun)$ & $a$ (AU) & $r$ (km) & $r_{atm}$ (km) & $r_p$ (km) \\ \hline 
 Mercury & 6023600.0 & 0.387 &  2437.6  &   0 &  2437.6  \\
 Venus   & 408523.71 & 0.723 &  6051.84 & 110 &  6161.84 \\
 Mars    & 3098708   & 1.523 &  3389.92 & 50  &  3439.92 \\
 Jupiter & 1047.3    & 5.202 & 71492    & 200 & 71692    \\
 Saturn  & 3497.898  & 9.514 & 60268    & 600 & 60868    \\
 Uranus  & 22902.98  &19.22  & 25559    & 350 & 25909    \\
 Neptune & 19412.24  &30.185 & 24764    & 250 & 25014    \\
\end{tabular}
\end{center}
\caption{Planetary data used in this study. Masses are from
  \cite{sta98}. The planetary radius $r_p$ used is the sum of the
  usual planetary radius $r$ plus the height of the atmosphere
  $r_{atm}$ to a density of $10^{-6}$~kg~m$^{-3}$ and is derived from
  \cite{lodfeg98}.}
\label{ta:planetarydata}
\end{table}

If the particle does not collide with the planet, scattering proceeds.
A matrix $M$ effecting a rotation by an angle $\gamma$ around the axis
defined by a unit vector along the pole $\vec{u} = (u_x,u_y,u_z) =
\vec{U}/|\vec{U}|$ and following the right-hand rule is given by
\begin{equation}
M = 
\left[ \begin{array}{ccc}
c + (1-c) u_x^2      & (1-c)u_yu_x - s u_z & (1-c)u_z u_x + s u_y \\
(1-c)u_x u_y + s u_z & c + (1-c) u_y^2     & (1-c)u_z u_y - s u_x \\
(1-c)u_x u_z - s u_y & (1-c)u_y u_z + s u_x & c + (1-c) u_z^2     \\
\end{array} \right]
\end{equation}
where $c = \cos \gamma$ and $s = \sin \gamma$, though alternate
formulations exist \citep{fraduncol55,hil01}. This matrix is used to
rotate the initial planetocentric velocity through an angle $\gamma$
to produce the post-scatter planetocentric velocity
$\vec{V_f}$.

The planet's velocity $\vec{v}_p$ is then added back to $\vec{V_f}$ to
yield the final heliocentric velocity $\vec{v_f}$. The particle is
taken to leave the Hill sphere of the planet at a position $R_f$
corresponding to a rotation of $R_i$ around $\vec{U}$ by an angle $\pi
+ \gamma$.

After the final heliocentric position and velocity are calculated, the
meteoroid is then assumed to follow a two-body orbit around the Sun.
The usual orbital elements can be calculated, and the bound or unbound
nature of the orbit assessed as well as whether or not it will reach
the Earth's orbit and at what velocity.

This technique for calculating the scattering of a meteoroid by a
planet is approximate, and breaks down where the motion of the planet
deviates appreciably from a straight line (\ie where scattering takes
a long time, or equivalently where the approach velocity is
low). However this is a minor issue for this study, as we will see
that hyperbolic scattering occurs almost exclusively for meteoroids
that are already moving near the hyperbolic limit and thus the
encounters are rapid and the straight-line approximation to the
planet's velocity is valid.

\subsection{The flux of hyperbolics at Earth}

The scattering algorithm of section~\ref{scattering}
provides some information on whether or not a given planet can produce
hyperbolic meteoroids at the Earth. What one would really like to
know is the absolute flux of such meteoroids at Earth to determine
whether they arrive in significant numbers, especially with respect
to the flux of interstellar particles.

The primary difficulty here lies in determining the flux of meteoroids
as a function of velocity and approach direction near the scattering
planet. Given the highly heterogeneous flux of meteoroids at the
Earth, we would expect a similarly complex meteoroid environment
around the other planets. Since no detailed measurements of the
meteoroid environments near other planets currently exist, we will
address the question first assuming one of two idealized meteoroid
environments, and then third modelled environment.

\begin{enumerate}
\item Each planet's Hill sphere is bombarded uniformly by meteoroids
  from all directions at all speeds compatible with the meteoroids
  being initially bound to the Sun.
\item Each planet's Hill sphere is bombarded by meteoroids travelling
  at all bound speeds but with their velocity vectors parallel to the
  planet's orbital plane. The meteoroids travel parallel to but not
  necessarily in the planet's orbital plane: they may enter the Hill
  sphere at positions above or below it. This mimics a scenario where
  meteoroids are concentrated in the ecliptic plane.
\item The sporadic meteor model at Earth of \cite{wievaucam09} is used
  to extract model meteoroid environments for the planets.
\end{enumerate}

In the first case, $10^{10}$ initial conditions are selected each with
an initial heliocentric speed drawn from a uniform random distribution
ranging from $0$ to $\sqrt{2} v_p$, where we note that our upper limit
is the maximum heliocentric speed that a meteoroid bound to the Solar
System can have. The initial approach directions are drawn uniformly
on a sphere. The meteoroid enters the Hill sphere at a point
($x_{tp}$, $y_{tp}$) on the target plane where each coordinate is drawn
from a uniform random distribution ranging from $+H$ to $-H$.  After
scattering, we tabulate those meteoroids which are unbound
(hyperbolic) relative to the Sun and in particular those which reach
the Earth.

A hyperbolic meteoroid scattered off an outer planet is considered to
be detectable at the Earth if the meteoroid is travelling inward on
the post-scatter leg of its orbit, has a perihelion $q < 1.1$~AU and
either $i$) its inclination to the ecliptic plane is less than
$3\degree$ or $ii$) it has a node within 0.1~AU of the Earth's
orbit. Hyperbolic meteoroids scattering from the inner planets
inevitably reach 1 AU and so only the conditions on the inclination
and/or node are applied to them. Note that we ignore any possible
gravitational focusing due to the Earth's gravity, which would be
small for fast-moving particles anyway.

\begin{figure}
\epsscale{1.0}
\plottwo{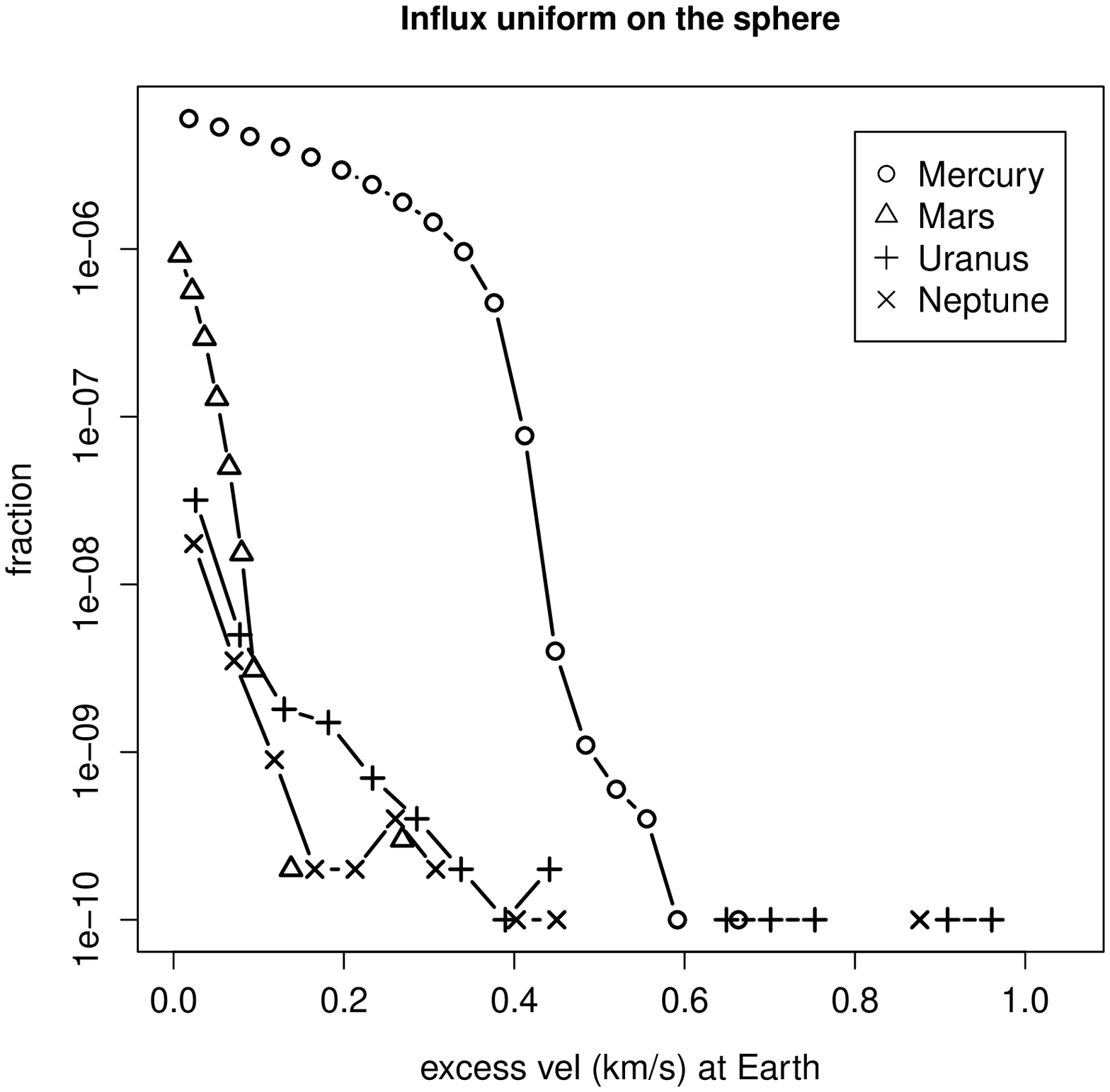}{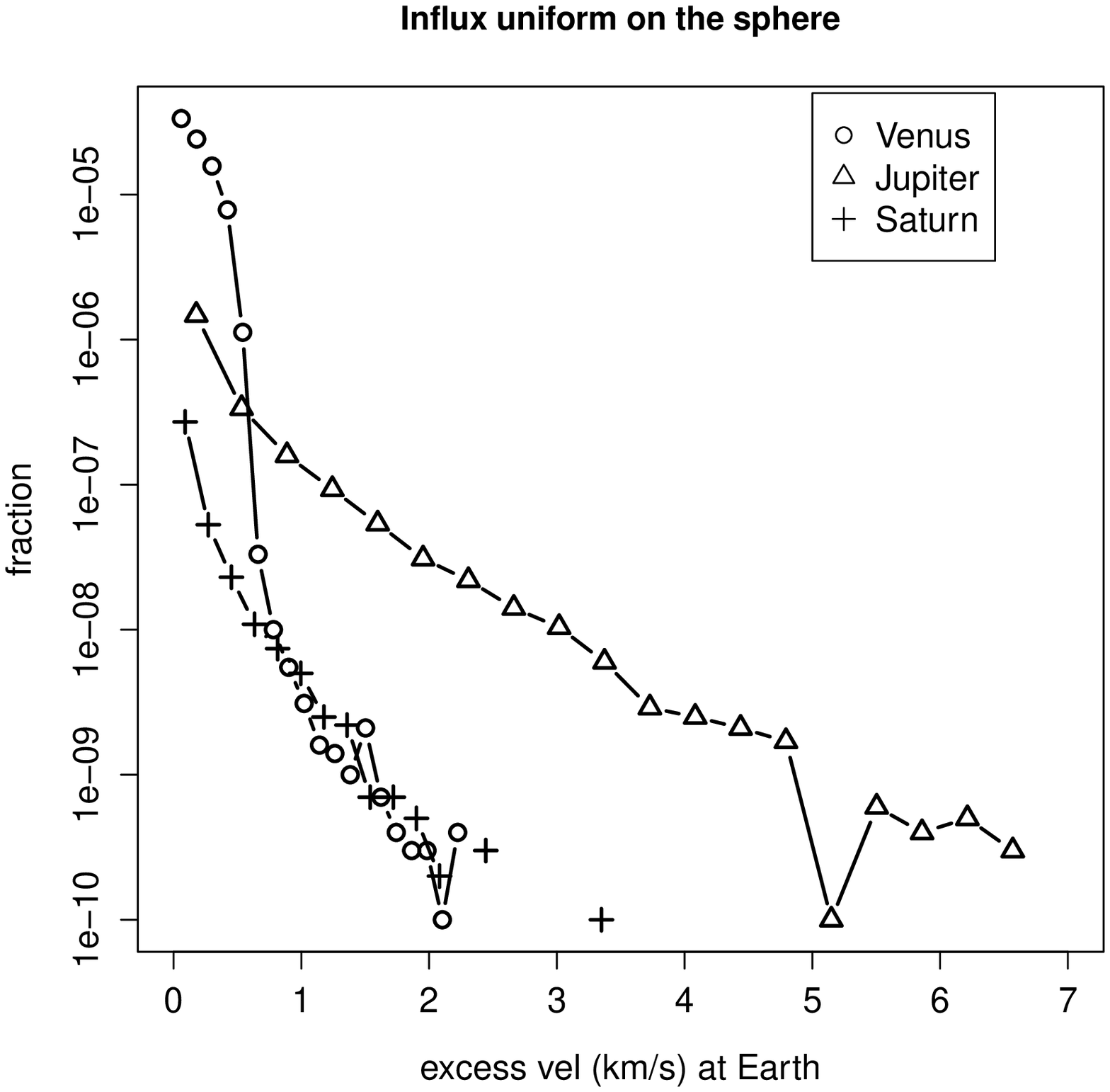}
\caption{The excess velocity with which a meteoroid reaches the
  Earth's orbit in the case of the planet being uniformly bombarded by
  particles from all directions.}
\label{fi:excess1}
\end{figure}

The results for the first case are shown in
Figure~\ref{fi:excess1}. The excess velocity $\vex$ is the velocity
above the hyperbolic limit {\it as it would be measured at the Earth};
Venus and Saturn are both able to produce velocities at Earth reaching
$\vex\approx 2$~km~s$^{-1}$. Jupiter creates the broadest distribution
of $\vex$, exceeding 6~km~s$^{-1}$. We see the result first reported
by \cite{wie11} that excess velocities comparable to those expected of
interstellar meteoroids (\ie a few km~s$^{-1}$) can be generated by
scattering off the planets.  Mars, Uranus and Neptune contribute much
less and at excess velocities mostly below 100 m~s$^{-1}$. Mercury
contributes much more than these last three, though the distribution
drops off sharply at $\vex \approx 400$~m~s$^{-1}$.

\begin{figure}
\epsscale{1.0}
\plottwo{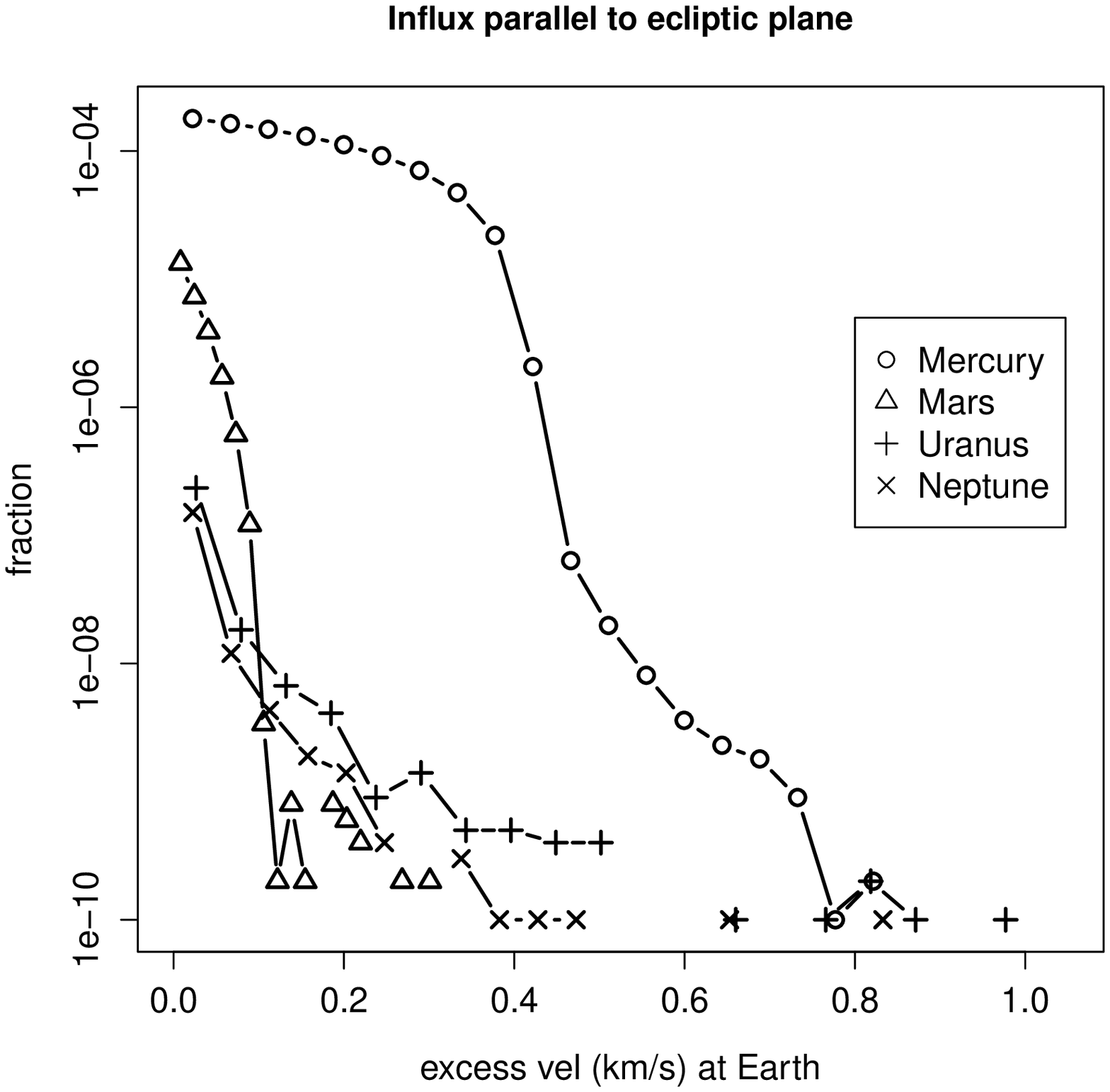}{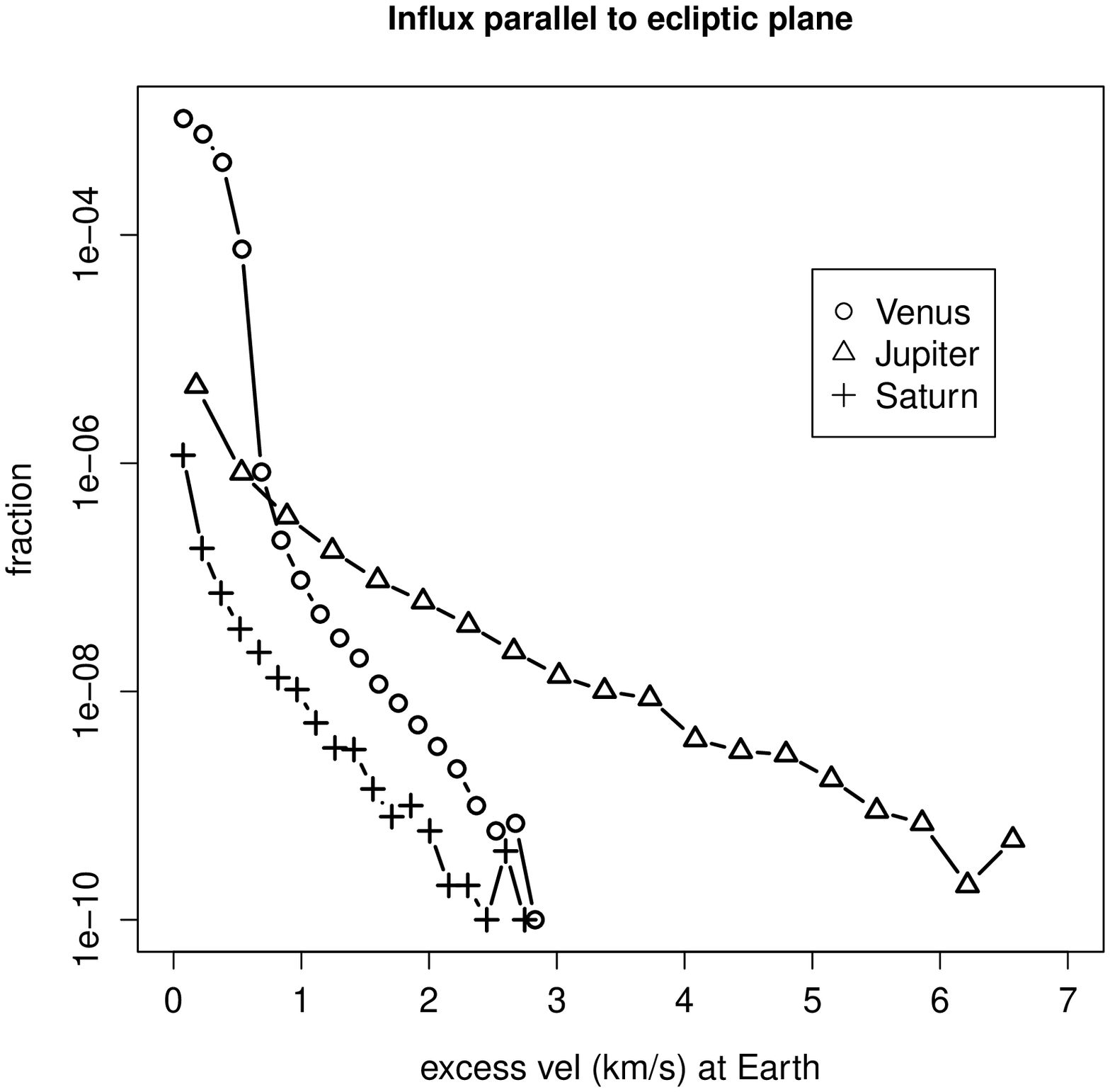}
\caption{The excess velocity with which
a meteoroid reaches the Earth's orbit in the case of the planet being
bombarded by particles travelling parallel to the planet's orbital plane.}
\label{fi:excess2}
\end{figure}

The second case, where the particle paths are concentrated in the
ecliptic is shown in Figure~\ref{fi:excess2}. The planets scatter more
efficiently in this case and the numbers at Earth are increased, but
the same qualitative trends that were noted for case 1 apply
here. Venus and Mercury are the most prolific produces of unbound
particles at the Earth though these more almost exclusively at $\vex <
1$~km~s$^{-1}$. while the largest $\vex$ are produced by Jupiter,
again reaching over 6 km~s$^{-1}$.

Note that the maximum $\vex$ described above do not represent true
physical maxima to these quantities, but reflect in part the finite
number of Monte Carlo trials. The largest excess velocities are
associated with meteoroids which pass very near to the surface (or top
of the atmosphere) of the planets, a very narrow window or 'keyhole'
which may not be well sampled by our Monte Carlo simulations. Thus the
possibility of even higher excess velocities than those described here
certainly exists.

To examine the size of the `keyholes' involved in some cases, plots
were created of the target plane, that is the $x_{tp}$--$y_{tp}$
plane. An example is shown for the largest $\vex$ seen at the Earth
via scattering from Uranus, which has $v_i = 1.345340 V_c$, $\phi=
77.\degree 833635$ and $\psi= 0\degree$. Given that $v_i$ is not
particularly close to $\sqrt{2}  V_c \approx 1.4142 V_c$, a strong
scattering to Earth might seem unlikely (as indeed it is). But a
check of the target plane shows that this scattering is correctly
calculated.

\begin{figure}
\epsscale{0.9}
\plotone{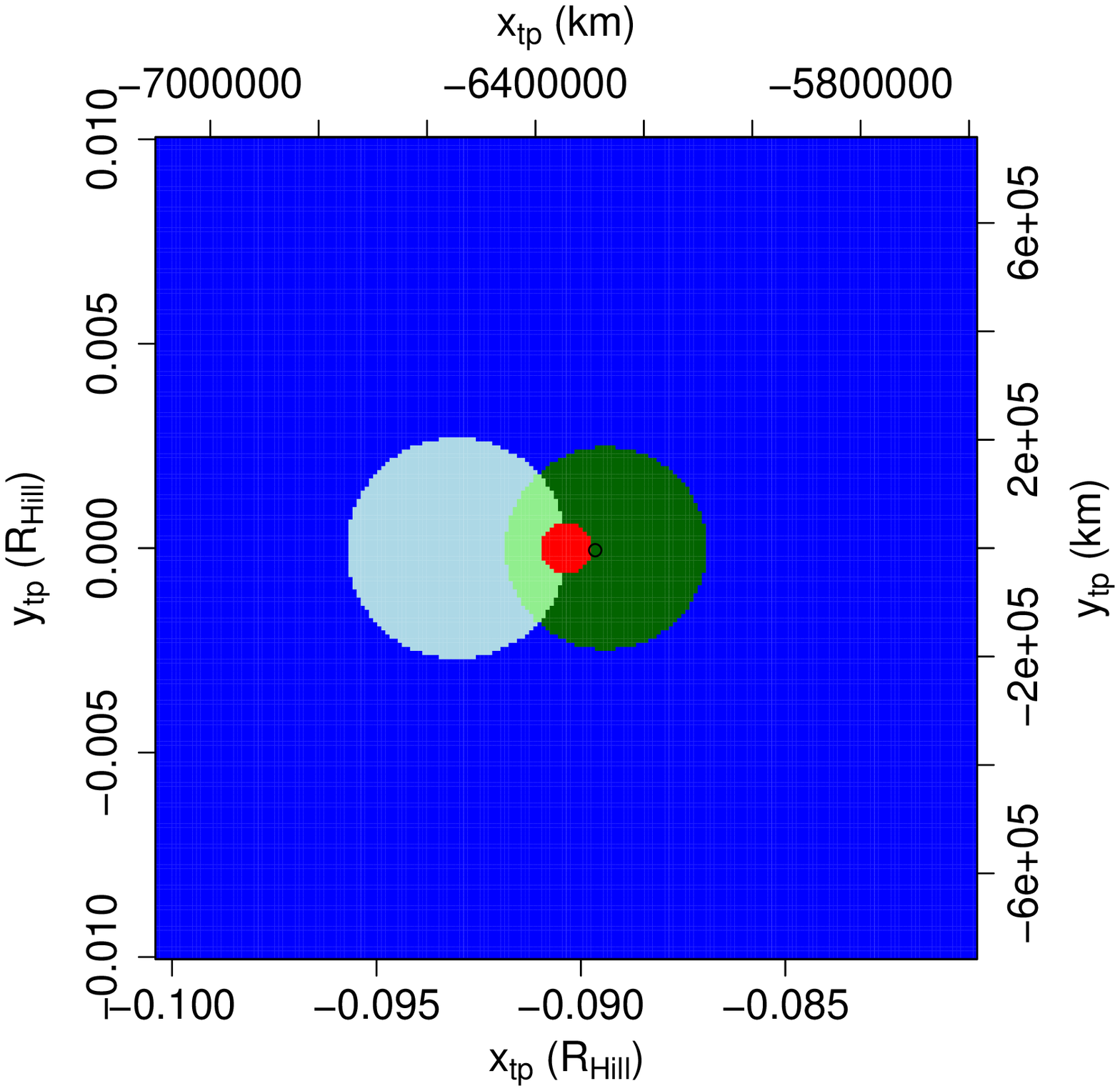}
\caption{A portion of the target plane near the largest $\vex$
  simulated for Uranus. The outcomes are coded by colour; the reader
  may wish to consult the on-line figures. Red indicates collision
  with the planet; white indicates that that the initial conditions of
  the particle recede from the planet or miss the Hill sphere; dark
  and light blue indicate bound heliocentric orbits moving inwards and
  outwards respectively; dark and light green indicate unbound orbits
  moving inwards or outwards respectively but that do not intercept
  Earth's orbit; yellow and orange indicate post-scattering orbits
  that are unbound and intercept the Earth's orbit with $\vex$ less
  than or greater than 1~km~s$^{-1}$ respectively. Not all colours
  appear on this plot.}
\label{fi:uranusTPa}
\end{figure}

\begin{figure}
\epsscale{1.0}
\plotone{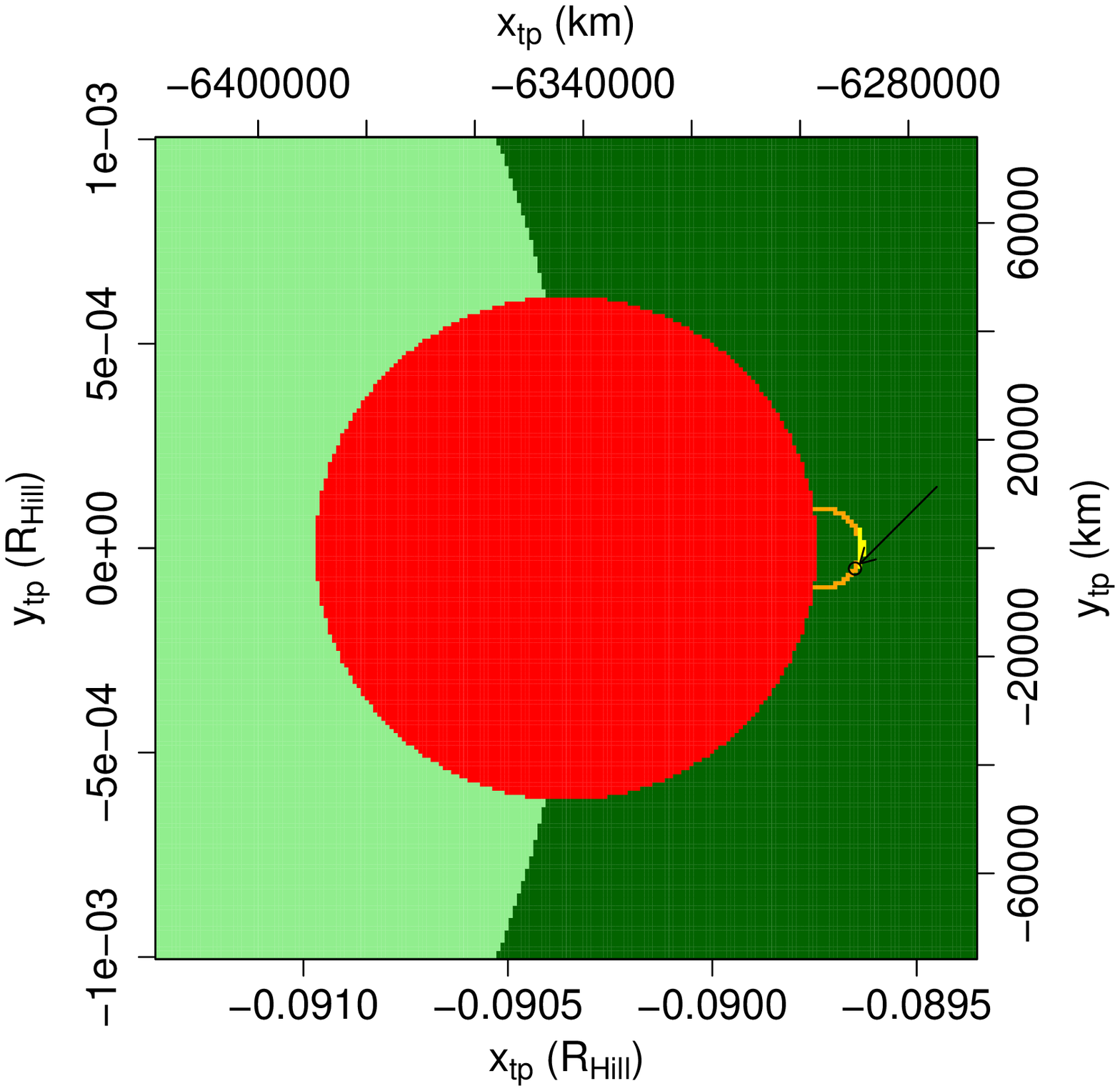}
\caption{A portion of the target plane of the largest $\vex$ simulated
  for Uranus. The $x$ and $y$ target plane coordinates of the
  meteoroid with the largest excess velocity seen at Earth from this
  planet in the Monte Carlo simulations is indicated by an arrow. See
  Figure~\ref{fi:uranusTPa} for the colour coding.}
\label{fi:uranusTPb}
\end{figure}

Figure~\ref{fi:uranusTPa} shows the part of the target plane where
this particular case occurs, revealing that it is the result of a
near-collision with the planet (red region). However the resolution of
Figure~\ref{fi:uranusTPa} is insufficient to reveal the true outcome.
Magnifying the portion of the target plane in question
(Figure~\ref{fi:uranusTPb}), a small arc only a few thousand
kilometres across appears wherein hyperbolic velocities at Earth can
be obtained (yellow and orange region). The case is produced by a
small keyhole that is not evident in coarse-grained examinations of
the target plane.

In fact, though the case we are examining produces a $\vex \sim
1$~km~s$^{-1}$, the largest $\vex$ in the orange region shown in
Figure~\ref{fi:uranusTPb} reaches over 1.5 km~s$^{-1}$; increasing the
resolution by a factor of five in each dimension reveal cases of $\vex
> 1.6$~km~s$^{-1}$. The reason such values do not appear in our
earlier figures is that the number of trials in our Monte Carlo
simulations ($10^{10}$ per planet) is not sufficient to sample all
these tiny regions well. Further examination of this and many other
keyholes found would be necessary to determine the maximum $\vex$ that
could be produced, but this is not attempted here since we are more
interested in typical values that the maxima. However, the largest
$\vex$ values for each planet produced in our Monte Carlo simulations
are examined by means of target plane plots like those of
Figure~\ref{fi:uranusTPb} and all are found to be real high-$\vex$
scattering events and not numerical or other errors.

The computations involve a large number of orbits near the parabolic
limit ($e \sim 1$) and so provide a data set which challenges most
orbital element calculation algorithms. We have avoided the use of the
orbital elements in the scattering calculations, expressing the
transformations in terms of rotations and additions of velocity
vectors to avoid such issues. A second potential numerical issue
involves scatterings with very small impact parameters $B \sim 0$ or
planetocentric velocities $|\vec{V}_i|^2 \sim 0$. In these cases,
$\gamma$ (Eq.~\ref{eq:gamma}) is ill-defined. However, these
correspond to collisions with the planet rather than scattering events
and are treated correctly here (that is, they are discarded).

An analytical determination of the the maximum $V_f$ possible (if the
planet were a point mass) is $V_f = V_i + 2 v_p$, a gain of twice the
orbital velocity of the planet.  However this provides only a loose
upper limit to our study of $\vex$. The restrictions due to the
planet's finite size together with the necessity of a post-scattering
intersection with the Earth's orbit complicate an analytical
approach. Thus we have not attempted to calculate the maximum possible
speed that a planet-scattered meteoroid might have at the Earth: the
intent of this study is only to present the general characteristics of
the post-scattering population of hyperbolic meteoroids, and to show
that interstellar speeds are possible.

\section{Expected fluxes of scattered sporadic meteors at Earth}

Cases 1 and 2 provide some idea of the range of possible scattering
events but the actual component of scattered hyperbolics in our Solar
System depends strongly on the meteoroid environment of the scatterer.

To address the question of the expected flux of hyperbolics at Earth,
we use a theoretical model for the sporadic meteoroid flux at the
planets as the input to our scattering model. \cite{wievaucam09}
numerically simulated dust released from a variety of comets and
asteroids and compared the results with radar observations of meteors
at the Earth. This allowed the primary contributors to the sporadic
meteors at the Earth to be determined and provides a broad-strokes
description of our planet's meteoroid environment. Using the data from
the same dynamical simulations, we are able to extract the meteoroid
environments for any of the planets and feed them into the scattering
model.

\begin{figure}
\epsscale{1.0}
\plotone{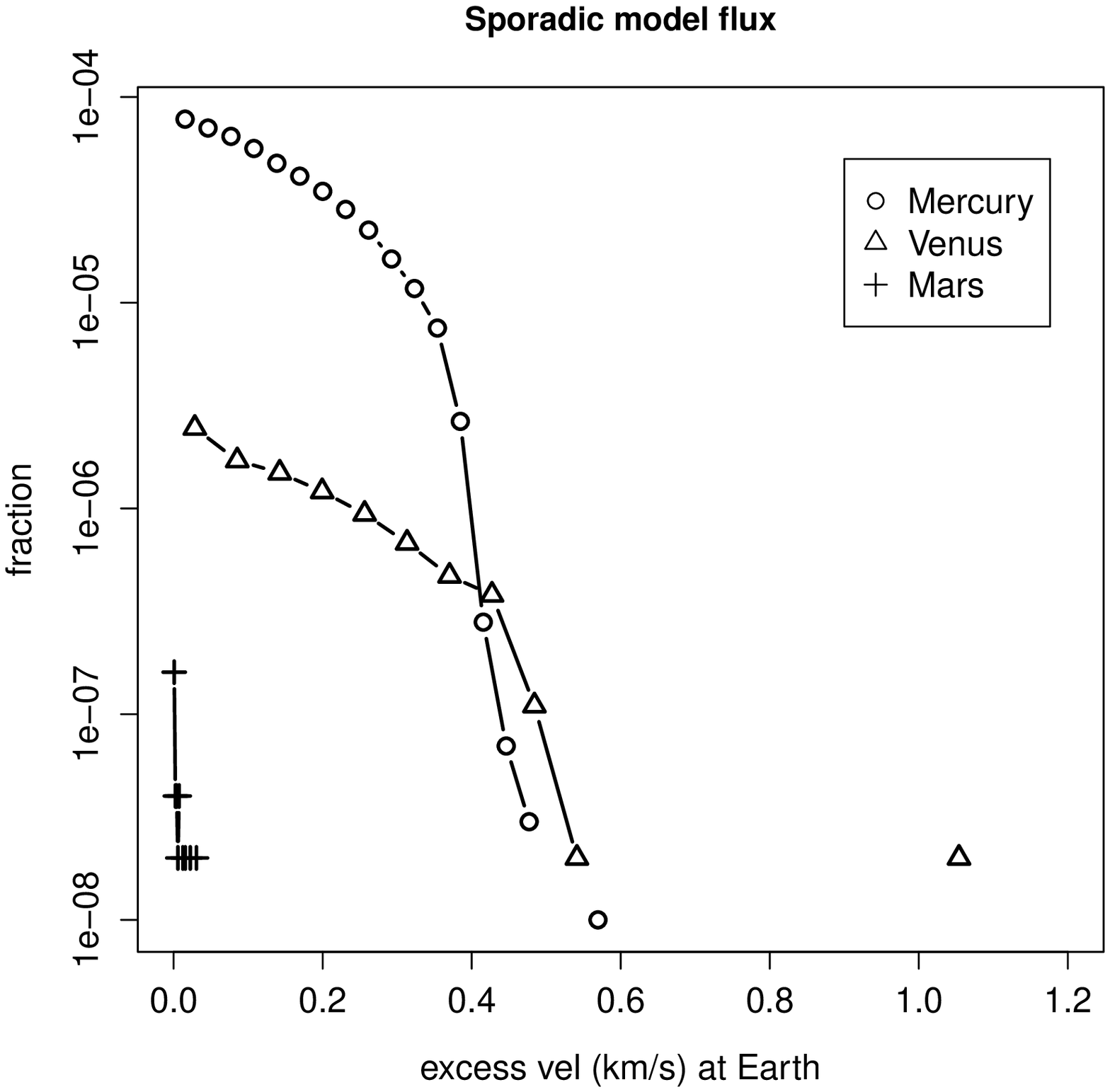}
\caption{The excess velocity with which a meteoroid reaches the
  Earth's orbit in the case of the planet being bombarded by a
  particle distribution derived from the sporadic model of
  \cite{wievaucam09}. Here the $y$-axis is the fraction of the total
  sporadic flux of meteoroids onto the planet's target plane that are
  accelerated to hyperbolic speeds.}
\label{fi:excess3}
\end{figure}

The results are presented in Figure~\ref{fi:excess3}. In this case,
only $10^8$ Monte Carlo trials were performed due to the more involved
calculations and hence slower execution speeds; each planet required
in excess of 10 CPU-days to complete.  Only the planets Mercury, Venus
and Mars produced any hyperbolic meteors at the Earth. Mercury is
perhaps unexpectedly the largest contributor. Though not able to
scatter to high excess velocities, it is a relatively effective
scatterer (with a surface escape speed of 4.2 km/s) and its environment is
rich in nearly-unbound meteoroids. Mercury's maximum simulated $\vex$ was
$580$~m~s$^{-1}$ while the mean and median values were 125 and
110~m~s$^{-1}$. Venus' maximum $\vex \approx 1080$~m~s$^{-1}$, with a
mean and median of 160 and 130~m~s$^{-1}$ respectively. Mars
scattered Earth-intercepting hyperbolics only to a maximum of
$\vex\approx 30$~m~s$^{-1}$ with mean and median values of only a few
meters per second, scarcely detectable.

The outer planets do produce hyperbolics, and they scatter meteoroids
onto paths that eventually reach Earth, but none that fit into both
categories. This results in part from the much lower average meteor velocity
in the outer solar system than in the inner, though the geometry of
encounter also plays a role. Thus in our Solar System, the least
massive planets are the chief contributors to the high-velocity
scattered meteoroid population near Earth, though the excess
velocities do not reach values comparable to those expected of
interstellar meteoroids. We conclude that the sample of interstellar
meteors may be contaminated by the scattering of sporadic meteors at
excess velocities below 0.5~km/s but not at higher $\vex$.

Can we estimate the absolute flux of contaminating scattered
hyperbolics? By extracting the number of meteors intersecting each
planet's Hill sphere per unit time from the model of
\cite{wievaucam09}, and multiplying by the fraction scattered on to
Earth-crossing hyperbolic orbits, we can determine the number of such
hyperbolics created by each planet per unit time. Relatively large
numbers of such meteors are produced, but the number that reach the
Earth is reduced by a factor related to the probability that any individual
scattered meteor will encounter our planet as it passes its orbit.
Travelling near the escape velocity (42 km/s) as they pass 1
AU, such particles have a probability $p$ of encountering the Earth
given roughly by the time needed to cross a torus the width of the
Earth ($t_{cross} = 2 R_{Earth}/v)$ divided by Earth's orbital period
so that
\begin{equation}
p \sim \frac{2\cdot 6378\rm{km}}{42 \rm{km/s}\cdot  3.15 \times 10^7} \rm{s} \sim 10^{-5}
\end{equation}
The end result is that the average rate of Mercury-scattered
hyperbolics arriving at Earth is only $10^{-4}$ the usual sporadic
meteor background. This fraction drops to $2 \times 10^{-5}$ for Venus
and $5\times 10^{-6}$ for Mars. This implies that observational sample
sizes of $\sim 10^{5}$ meteors would be needed before significant
number would be observed and so we do not expect many yet to have been
seen. Because of the restrictions of the geometry of scattering to
Earth, these scattered hyperbolics do not arrive uniformly in time but
peak at intervals of the scatterer's synodic period, and so the rates
may vary by an order of magnitude or more with time.

\begin{figure}
\epsscale{1.0}
\plottwo{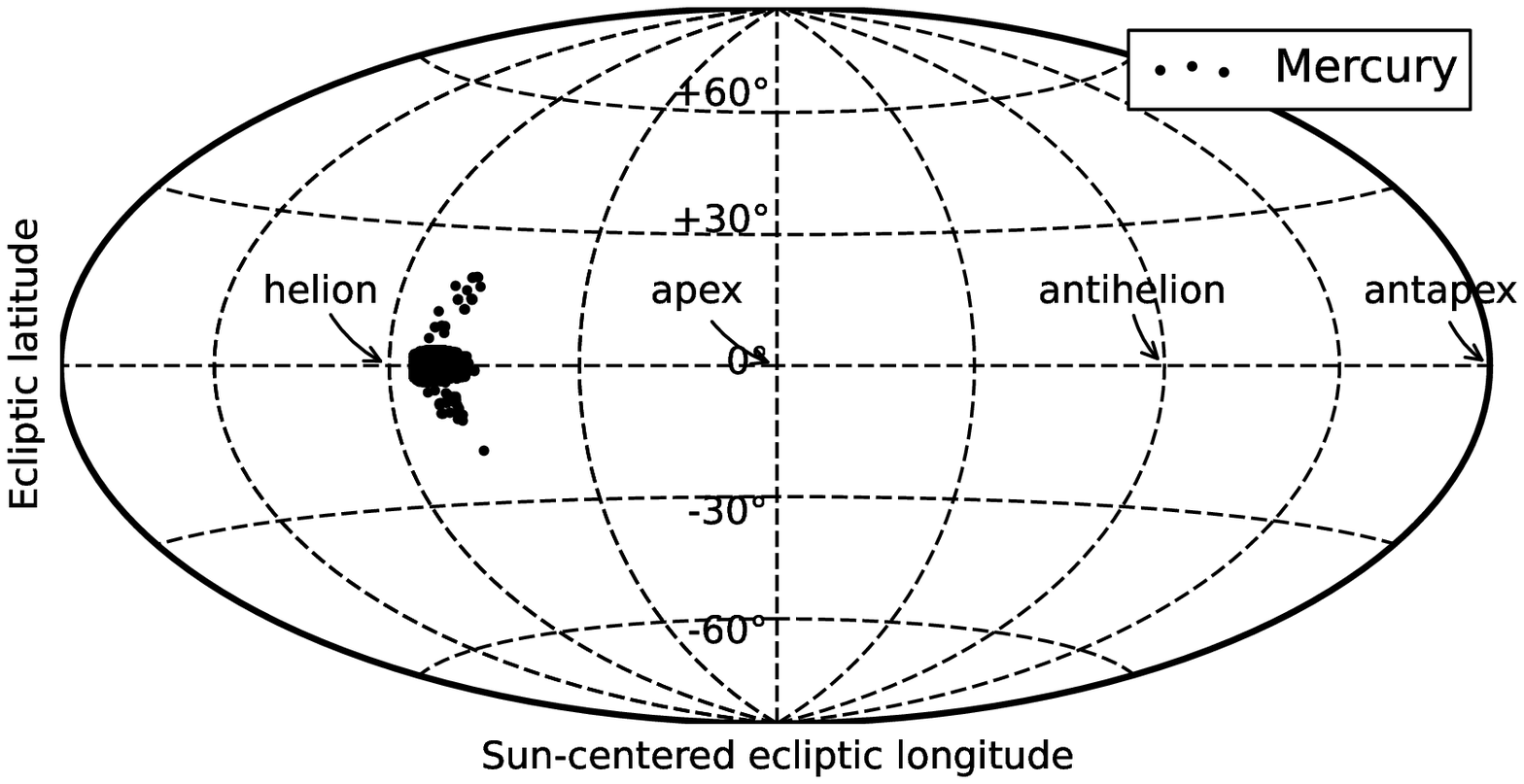}{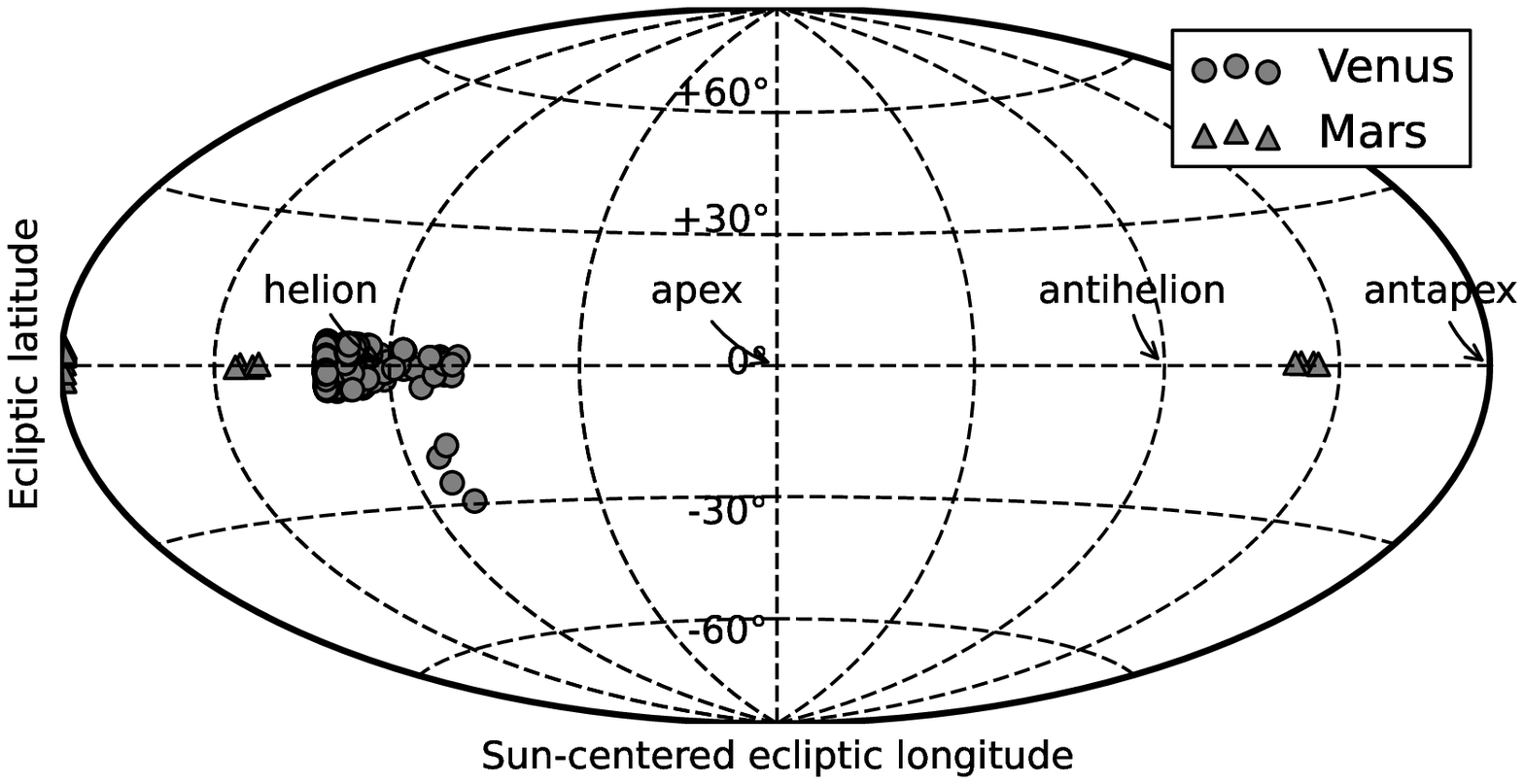}
\caption{The hyperbolic radiants at Earth produced by the planets in
a Sun-centered ecliptic coordinate system.}
\label{fi:hyperbolicradiants}
\end{figure}

\begin{figure}
\epsscale{1.0}
\plottwo{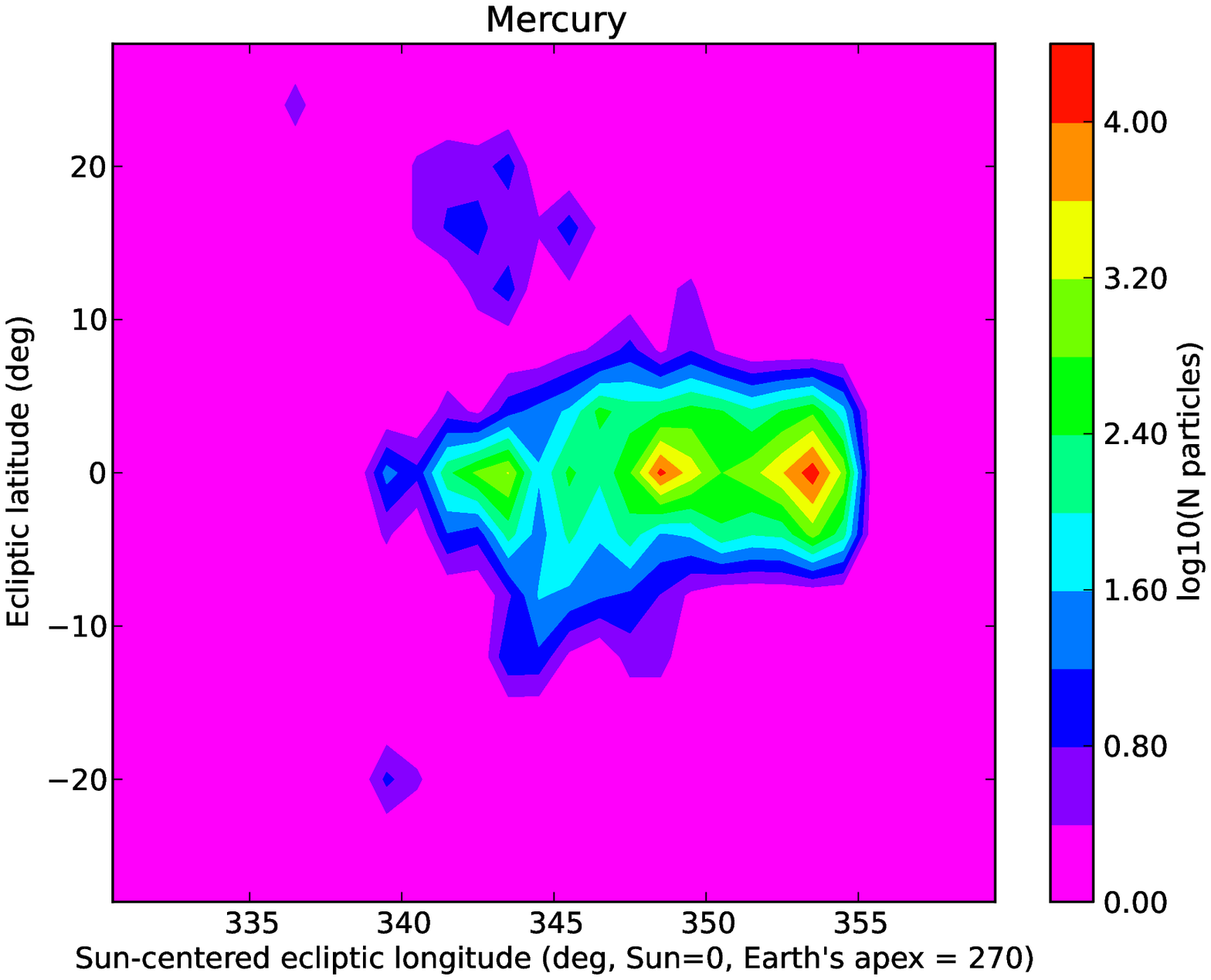}{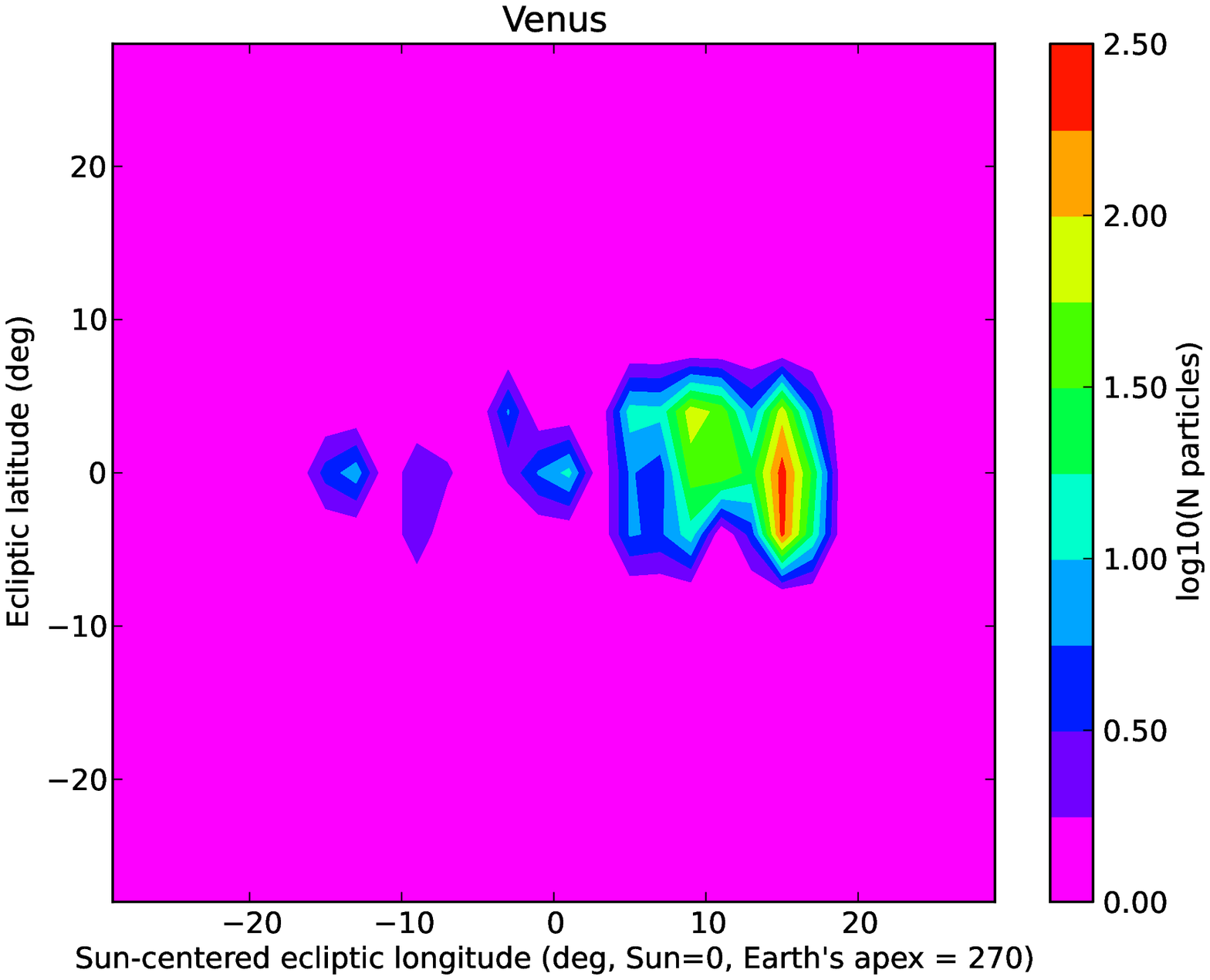}
\caption{Magnified view of  hyperbolic radiants at Earth produced by Mercury and Venus}
\label{fi:hyperbolicradiants-zoom}
\end{figure}

Though small in number, the arrival times and directions of these
scattered hyperbolics are correlated. The radiants observed at the
Earth (Figure~\ref{fi:hyperbolicradiants} and
\ref{fi:hyperbolicradiants-zoom}) are concentrated near the Sun for
the inner planets, and near the antapex for Mars. Thus there should be
'hyperbolic meteor showers' with distinct radiants which re-occur with
the synodic period of the scatterer: 116 days for Mercury, 584 days
for Venus, 778 days for Mars.

Accounting for the travel time of the particles from the scatterer to
Earth, we can estimate when in the synodic cycle of the scatterer that
the hyperbolic radiant should occur. The travel time from Earth to
Mercury for simulated hyperbolics is 20-50 days with peaks near 27 and
40 days; for Venus, 10-70 days with a peak near 33 days and for Mars,
40-100 days peaking at 68 days. From this we can estimate that a
hyperbolic shower should occur whenever Mercury is 30-60$\degree$
ahead of the Earth in its orbit, near its largest western
elongation. For Venus, the shower should occur when that planet is
within 10$\degree$ of inferior conjunction. For Mars, when it is
30$\degree$ past opposition, though there is considerable scatter for
this planet.  Searches for the appearance of such radiants should be
conducted near these times, which are readily available from
astronomical almanacs.

The radiants produced by the inner planets are located near the Sun
(see Figure~\ref{fi:hyperbolicradiants-zoom}) and thus will be buried in
the helion sporadic meteor source.  For Mercury, on the ecliptic at a
sun-centered longitude of 350$\degree$ with relative velocities (far
from Earth) of 33-40 km/s; and for Venus, on low latitude radiants
centered more or less on the Sun at relative velocities of 24-30
km/s. These radiants might not be easily observed optically because
they are in the daytime sky, but should be detectable by modern radar
meteor observatories. For Mars, the few hyperbolic meteoroids are
concentrated in the antapex direction (see
Figure~\ref{fi:hyperbolicradiants}) with relative velocities of 12-15
km/s.

Have any of these radiants already been seen? First, one might ask if
any reported hyperbolic meteors had radiants near the Sun or the
antapex. On the radar side, AMOR \citep{bag99} hyperbolics were
observed at a wide range of inclinations, while the hyperbolic showers
mentioned above would be concentrated at low inclinations. A CMOR
search \citep{werbro04} collected data from 2002-2004 and so may
contain some, but they do not present enough information to determine
whether that was the case. Arecibo data \citep{meijanmat02a} were
collected over 1997-1998 and could contain such detections: many of
them are at low ecliptic latitude. \cite{janmeimat01} indicate that
many of these originate in the Mercury-Venus zone, but there is no way
to determine for sure from the information as presented there.

As far as optical detections go, none of the most likely hyperbolic
meteors of \cite{jonsar85} have inclinations in the ecliptic. The two
optical meteors of \cite{hawwoo97} also did not have low inclinations.
\cite{muswerbro12} do present a plot of their image-intensified camera
radiants in Sun-centered ecliptic coordinates but none of their 22
possible interstellar meteoroids have radiants near the antapex or the
Sun.

The current largest collection of meteor orbits, from the Canadian
Meteor Orbit Radar (CMOR) in London Canada, has several million orbits
\citep{jonbroell05} so we might expect to already have such detections
among them. Though at least part of the data has been searched for
interstellar meteors, no such hyperbolic radiants have been reported.
However, their analysis was probably not designed to reveal the
presence of radiants as described here.  The examination of CMOR data
for interstellar meteors by \cite{werbro04} examined 1556384
individual detections and found fewer than 12 and 40 were 3 and
$2\sigma$ above the hyperbolic limit respectively. Our theoretical
result here might have predicted $\sim100$ hyperbolics from Mercury
alone in this sample. However, the stringent detection limits of
\cite{werbro04} required that the velocity errors, about 15\% in the
$2\sigma$ case, be small enough that any excess velocity could be
clearly assigned to the detection itself and not to observational
error.  Since a typical speed for their sample was 56-68~km~s$^{-1}$,
a 15\% error amounts to several km~s$^{-1}$, and would far exceed the
small excess velocity predicted to be produced by Mercury, Venus or
Mars. The CMOR radar may not currently have the accuracy necessary to
detect planet-scattered hyperbolics as unambiguously unbound, but a
search of the database specifically for such meteors is planned for
future work.

We conclude that these hyperbolic meteor showers have not yet been
detected; nonetheless we look forward to ever-improving detection
methods and new sensors with a hope that soon the existence of
meteoroids arriving at Earth directly from near-planet environments
can be confirmed or falsified.

\section{Scattered shower meteors at Earth}

One aspect that the model above does not consider is the contribution
from meteoroid streams, as the sporadic model of \cite{wievaucam09}
specifically excludes showers. A well-populated debris stream that
crosses a planet orbit would provide a fresh set of particles to
scatter each time the planet passed through it and so could be an
important if erratic source. To include the possible contributions
from showers, we examine those comets which have nodes near a planet,
and ask if meteoroids travelling on orbits like that of their parent
can be scattered onto hyperbolic orbits that take them to Earth.

All comets in the JPL comet catalogue downloaded from JPL Horizons on
Feb 25 2013 were examined for nodes within a Hill radius of one of the
seven planets (excluding Earth). To this list, the asteroidal parents
of two well-known streams 3200 Phaethon and 2003~EH$_1$ were added for
completeness.  If a node near a planet was found, then a
simulation is run where the entire Hill sphere is populated with
meteoroids travelling parallel to the trajectory that the parent would
as it passed the planet in question's orbit, and the results of the
gravitational scattering is analyzed. This approximates the case of
a meteoroid stream produced by the parent being scattered by the planet. 

Periodic comets are likely to have substantial dust trails associated
with them, but few are found to produce significant hyperbolics. An
exception is 2P/Encke. Dust from comet Encke is scattered quite
efficiently by Mercury. Though the post-scatter trajectories do not
take them to Earth, scattered particles reach eccentricities above 1.1
corresponding to $\vex=3$~km/s at 1~AU, and thus 2P may be a source of
hyperbolic meteoroids at other locations in our Solar System. The only
other periodic comet found to produce hyperbolic meteoroids (though
they don't reach Earth) is 177P/Barnard: scattering by Mars produces
eccentricities up to 1.02 in our simulations.  Jupiter-family comets
are abundant and do often have a node near Jupiter but do not appear
here, because the scattering geometry as well as the low relative
speeds are not favourable to producing hyperbolics.

Appreciable hyperbolic meteoroids could be produced by comets on
nearly-unbound orbits.  In practice, few of these produce meteoroids
at Earth's orbit but many could contribute to a broader population
with our Solar System. About one-third of the long-period comets with
nodes near one of the outer planets have a geometry favourable to
producing hyperbolic meteoroids of some description, though most of 
these don't reach the Earth.

In the case of a long-period comet, the assumption of a well-populated
meteoroid stream in the vicinity of the parent's orbit is unjustified.
As a result, further checks are needed to determine if the parent
comet's trail did indeed pass near the scattering planet in question,
and whether or not the Earth would have been in the correct location
to intercept the scattered meteoroids. The necessity of both the
scattering planet and the Earth being in the correct locations in
their orbit in order that a hyperbolic meteoroid be detectable at the
Earth considerably reduces the number of cases of interest. Overall
this process is not likely to contribute significantly to the flux of
hyperbolic meteoroids at the Earth except perhaps during brief
intervals. A few cases are discussed below.

Comet 1976/D1 Bradfield passed near Jupiter in 1974 as it approached
the Sun. Just above 4\% of the target plane scattered by this comet
could have produced hyperbolic meteoroids at Earth's orbit with $\vex$
up to 0.6~km/s. However the encounter conditions were such that the
comet arrived at Jupiter's orbit just after the planet had passed, and
so no scattering could have taken place.

Almost 90\% of the target planet Comet 1991 T2 Shoemaker-Levy with
Saturn would have produced hyperbolic meteoroids at Earth's orbit with
$\vex \las 0.3$~km/s. However that planet would not have passed
through the debris stream from this long-period comet until 2000
(seven years after the parent had passed that location) and thus it is
unlikely that there were any meteoroids present to scatter.

\begin{figure}
\epsscale{0.9}
\plotone{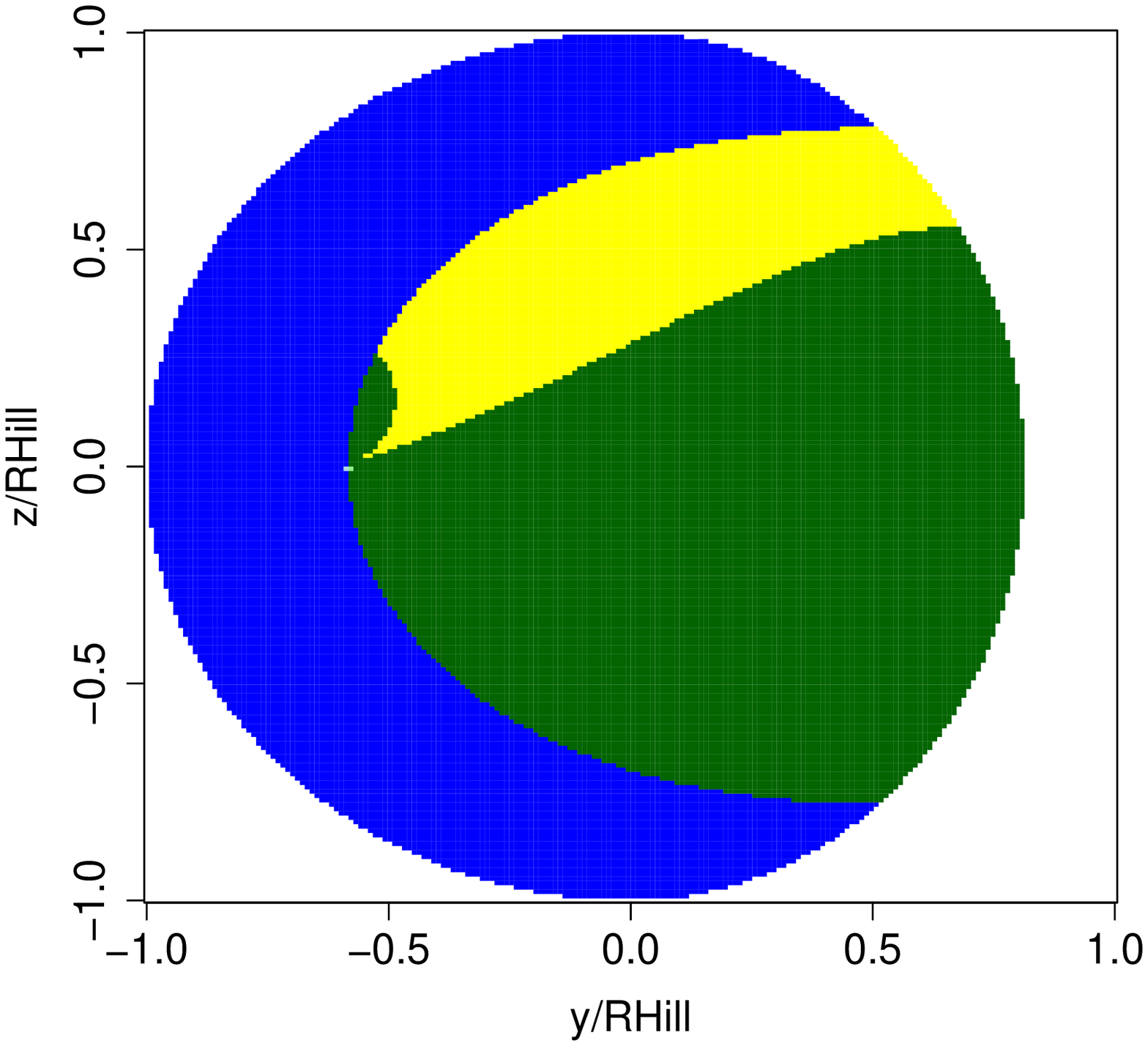}
\caption{The target plane of the encounter of C/1995 O1 Hale-Bopp with
  Jupiter. The outcomes are coded by colour; the reader may wish to
  consult the on-line figures. Yellow indicates unbound orbits moving
  inwards and intercepting Earth's orbit; dark green are inward-moving
  hyperbolics which do not reach our planet's orbit, and dark blue are
  inward-moving bound orbits. See the caption of
  Figure~\ref{fi:uranusTPa} for more details.}
\label{fi:HaleBoppTP}
\end{figure}

Perhaps the most interesting case is that of comet Hale-Bopp. On its
pre-perihelion leg in 1996, comet 1995~O1 Hale-Bopp passed just ahead
of Jupiter.  In this case, the possibility of the planet encountering
some portion of the the meteoroid stream is substantial. Fifteen
percent of the target plane would have scattered hyperbolic meteoroids
to Earth's orbit, and at $\vex$ of up to 1~km/s (see
Figure~\ref{fi:HaleBoppTP}).  To determine whether or not scattered
meteoroids would have been directed to Earth would require a full
investigation of this event along with numerical simulations of the
dust trail and its interactions with Jupiter. While very interesting,
this is not addressed here. In particular, a careful calculation of
the time for the scattered particles to travel to the Earth's orbit
would be required to determine whether or not such meteors might have
reached our planet and been observed.  Though the Earth did not pass
through the debris stream from this spectacular comet, it may have
nonetheless have encountered particles originating from
this parent but arriving via a non-traditional route.

Given the absence of automated meteor observing systems at the time of
Hale-Bopp's passage, it is unlikely that any such meteors would have
been recorded, though the European Fireball Network
\citep{obemolhei98} and the Advanced Meteor Orbit Radar (AMOR)
\citep{bagbenste94} were operating at this time. Note that
interstellar meteors reported by AMOR \citep{bag00} were collected in
the 1995-1998 time frame and so may contain some contamination from
this source.  The hyperbolic meteors reported by \cite{hawwoo97} were
observed in June 1995 --before Hale-Bopp passed Jupiter-- and so those
meteors could not have been associated with this scattering event.

\section{Conclusions}

The number of hyperbolic meteors observed at Earth that might be
produced by gravitational scattering by the planets is calculated:
overall the numbers are small, certainly compared to the background
rate of bound meteors. According to \cite{cambra11}, the sporadic flux
of video meteors at Earth is
$0.18\pm0.04$~meteoroids~km$^{-2}$~hr$^{-1}$ while for the interstellar
flux \cite{muswerbro12} give an upper limit of $2 \times
10^{-4}$~meteoroids~km$^{-2}$~hr$^{-1}$ at optical sizes, so current
limits from optical systems put the flux of interstellars at 1 in 1000
at most.  Our work here predicts 1 meteor in $10^4$ at Earth will be a
hyperbolic originating in a scattering event at Mercury. As a result,
the contamination of a sample of interstellar meteors in this size
range is expected to be at least at the 10\% level and possibly higher
and so needs to be addressed. The problem is mitigated by the fact
that the contamination is primarily at low $\vex$, much smaller than
those expected of true interstellars. At radar sizes, CMOR
\citep{werbro04} sees perhaps 1 in $10^5$ meteors as hyperbolic. Our
work here still predicts 1 in $10^4$ internally-generated hyperbolics
at these sizes, but again their excess velocities are too low to
constitute a significant source of confusion in current samples.

We conclude that hyperbolic particles can in principle be generated
wholly within our Solar System and at speeds which rivals those
expected of interstellar meteors. However, the properties of the
meteoroid environment of our planetary system are not conducive
scattering larger numbers of them onto Earth-intercepting orbits.
Though selecting a sample of presumed-interstellar meteors solely on
the basis of their heliocentric velocity is likely to produce a
substantially contaminated sample, the internally-generated
hyperbolics are relatively easy to account for, as their excess
velocities at the Earth are expected to be only about 100~m~s$^{-1}$
in most cases. Higher $\vex$ may occur in exceptional cases or when a
planet encounters a freshly-deposited debris stream from a comet.
However, in all such cases a sufficiently precise velocity
determination followed by a careful examination of the pre-atmospheric
trajectory can determine whether such a scattering event occurred.
Thus while the search for interstellar meteors is complicated by
planetary scattering, continuing improvements in detection methods
mean that the phenomenon is not likely to prove a substantial obstacle
to the study of interstellar meteoroids.

\acknowledgments{This research was supported in part by the Natural
  Sciences and Engineering Research Council of Canada and NASA's
  Meteoroid Environment Office.}

\bibliographystyle{icarusbib}
\bibliography{Wiegert}

\end{document}